\documentclass[twocolumn,showpacs,preprintnumbers,amsmath,amssymb,pra,superscriptaddress]{revtex4}

\usepackage{graphics}
\usepackage{epsfig}
\usepackage{bm}

\def\jpb{J.\ Phys.\ B: At.\ Mol.\ Opt.\ Phys.\ }

\def\beq{\begin{equation}}
\def\eeq{\end{equation}}

\def\eulere{\mathrm{e}}

\def\reff#1{(\ref{#1})}

\def\Rg{R_\mathrm{grid}}
\def\Up{U_\mathrm{p}}

\def\VI{V_\mathrm{I}}

\def\omegaMie{\omega_\mathrm{Mie}}
\def\omegalaser{\omega_\mathrm{l}}
\def\omegap{\omega_\mathrm{p}}

\def\Vxc{V_\mathrm{xc}}

\def\VH{V_\mathrm{H}}
\def\Ip{I_\mathrm{p}}

\def\N3d{N_\mathrm{3D}}

\def\rs{r_\mathrm{s}}
\def\ri{R_\mathrm{i}}
\def\ro{R_\mathrm{o}}

\def\vekt#1{\bm{#1}}

\def\vektr{\vekt{r}}

\def\vecK{\vekt{K}}
\def\vece{\vekt{e}}
\def\vecp{\vekt{p}}
\def\vecr{\vekt{r}}
\def\vecX{\vekt{X}}

\def\vekte{\vekt{e}}
\def\vektE{\vekt{E}}

\def\vektA{\vekt{A}}

\def\vektp{\vekt{p}}

\def\Edach{\hat{E}}

\def\zhat{\hat{z}}
\def\Ehat{\Edach}

\def\Ahat{\hat{A}}

\def\energy{{\cal{E}}}

\def\pabl#1#2{\frac{\partial #1}{\partial #2}}
\def\bra#1{\langle #1 \vert}
\def\ket#1{| #1 \rangle}
\def\braket#1#2{\langle #1 | #2 \rangle}

\def\imagi{\mathrm{i}}
\def\diff{\,\mbox{\rm d}}

\begin{document}
%\draft

\title{Recollision-induced plasmon excitation in strong laser fields}
\date{\today}
\author{M.\ Ruggenthaler}
\affiliation{Max-Planck-Institut f\"ur Kernphysik, Postfach 103980, 69029 Heidelberg, Germany}
\author{S.V.\ Popruzhenko}
\affiliation{Max-Planck-Institut f\"ur Kernphysik, Postfach 103980,
69029 Heidelberg, Germany}
\affiliation{Moscow State Engineering Physics Institute, Kashirskoe Shosse 31, 115409, Moscow, Russia}
\author{D.\ Bauer}
\affiliation{Max-Planck-Institut f\"ur Kernphysik, Postfach 103980, 69029 Heidelberg, Germany}

\begin{abstract}
Recolliding electrons are responsible for many of the interesting phenomena observed in the interaction of strong laser fields with atoms and molecules. We show that in multielectron targets such as C$_{60}$ a new important recollision pathway opens up: the returning electron may excite collective modes even if the laser frequency is far off-resonant. We formulate a simple analytical theory which predicts that the recollision-induced excitation of collective modes should dominate over the ``usual'' harmonic generation yield at 800~nm wavelength. In this case the  tomographic imaging of complex multielectron systems may be obscured. We employ a time-dependent density functional model of C$_{60}$ and show that with increasing laser wavelength the dynamics becomes more and more single active electron-like, suggesting that long wavelengths are to be preferred for imaging purposes. 
\end{abstract}


\maketitle

\section{Introduction}
A typical interaction scenario in strong field laser atom or molecule interaction involves three steps: (i) the removal of an electron from a target (ionization), (ii) motion of this electron in the continuum, and, possibly, (iii) a recollision with the ``parent'' atom or molecule if step (i) occurred at a time such that the laser field drives the electron back. The recollision in the third step is responsible for the plateaus in photoelectron and high harmonic spectra, and nonsequential multiple ionization, corresponding to the three pathways (i) scattering in the presence of a laser field, (ii) recombination and emission of a photon, and (iii) laser-induced collisional ionization (see, e.g., \cite{milo06,ago04,abecker05} for reviews).

Structural information about the target is encoded in both photoelectron and harmonics spectra. Hence, besides the potential of high order harmonic generation (HOHG) as an efficient source of short wavelength radiation and attosecond pulses \cite{ago04}, the so-called ``tomographic imaging'' of molecular orbitals (see \cite{itani04} and \cite{lein07} for a review) has attracted considerable attention. It is clear that whatever is ``imaged'' in this procedure is supposed to be representation-independent, i.e., should not depend on the basis in which one expands the multielectron wavefunction. This requirement is difficult to fulfill within the simple and commonly adopted single active electron approximation (SAE) \cite{patch07}. 

In this work we study the recollision dynamics and the emitted radiation for the case of the C$_{60}$ fullerene, which is an example for a multielectron system displaying collective modes and an interesting dynamics when exposed to fs laser pulses \cite{hertelreview,c60exp} (other such systems are, e.g., metal clusters or biomolecules). The laser frequency is kept well below the surface and volume plasmon frequency of  C$_{60}$ so that only the recolliding electron may excite the collective modes efficiently but not the laser itself. In the context of ``orbital imaging'' it is vital to know whether the structural information encoded in the HOHG spectra is ``contaminated'' by emission at collective frequencies.  In other words, we are interested in the {\em relative} efficiency of the collective response with respect to the ``standard'' harmonic generation.

The outline of the paper is as follows. In Sec.~\ref{mode} the C$_{60}$ jellium model we use in the time-dependent density functional theory (TDDFT) calculations is reviewed, and its collective modes are identified. In Sec.~\ref{resu} HOHG spectra are presented for three different wavelengths, ranging from the typical 800~nm up to 3508~nm. The transition from the linear to the nonlinear excitation regime is discussed, enhancements in the dipole spectra due to plasmon excitation are evidenced, and their origin is investigated. In Sec.~\ref{lewe} we compare the TDDFT results with the predictions of a simple, SAE Lewenstein-like model of HOHG from C$_{60}$.  Section~\ref{collmodel} is devoted to an analytical model which takes collective modes into account and enables us to predict the relative efficiency of harmonic emission due to recollision-induced plasmon excitation (RIPE) with respect to standard harmonic generation.
Finally we conclude in Sec.~\ref{concl}.

\section{Model} \label{mode}
\begin{figure}
\includegraphics[width=0.3\textwidth]{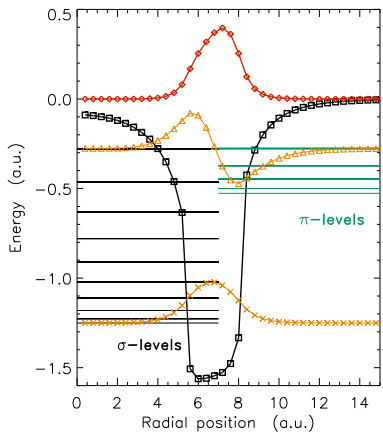}
\caption{(color online). Net KS potential (black, squares), total density (red, diamonds), wavefunctions of the lowest KS orbital and the HOMO (orange, crosses and triangles, respectively). The $\sigma$- and $\pi$-levels are indicated. Density and wavefunctions are scaled to fit into the plot.  \label{figpotential}}
\end{figure}

The   C$_{60}$ fullerene is modelled using density functional theory (DFT) employing a jellium potential for the ionic background of inner and outer radius $\ri$, $\ro$, respectively, \cite{puska,bauerC60}, i.e., 
\beq
V(r)= \left\{ \begin{array}{l} \displaystyle
 -\kappa\frac{3}{2} \left(\ro^2-\ri^2\right), \hspace{3cm} r\leq \ri \\
 \displaystyle  -\kappa \left(\frac{3}{2} \ro^2 - \left[\frac{r^2}{2}+\frac{\ri^3}{r}\right]\right) - V_0, \ \ \ri < r < \ro\\
 \displaystyle  -\kappa\frac{\ro^3-\ri^3}{r}, \hspace{3.7cm} r \geq \ro 
\end{array} 
\right.
\eeq 
where $\kappa=\rs^{-3}$, $\ri=5.3$, $\ro=8.1$, $\rs^{-3}=N/(\ro^3-\ri^3)$, $N=250$ Kohn-Sham (KS) electrons, and  $V_0=0.68$ (atomic units are used unless noted otherwise). The solution of the time-independent KS equation 
\beq 
\epsilon_j \ket{\psi_j} = (T+V+ \VH + \Vxc ) \ket{\psi_j}
\eeq 
yields the ground state configuration from which we start the propagation. Here, $\ket{\psi_j}$, $j=1\ldots N$ are the $N$ KS orbitals,  $\epsilon_j$ are the KS orbital energies, $T$ is the single-particle kinetic energy operator $p^2/2$,
\beq \VH=\int\diff^3 r'\, \frac{n(\vektr')}{\vert \vektr-\vektr'\vert}\eeq 
is the Hartree potential, 
\beq \Vxc(\vektr)=-\left[\frac{3 n(\vektr)}{\pi}\right]^{1/3}\eeq
is the exchange-correlation potential in  exchange-only local density approximation (LDA), and 
\beq n(\vektr)=\sum_j | \braket{\vektr}{\psi_j}\vert^2\eeq 
is the electron density.
The $N=250$ KS electrons lead to  a spin-neutral, closed-shell ground state of spherical symmetry. More precisely, we obtain $200$ $\sigma$-electrons (without node in the radial wavefunctions) and $50$  $\pi$-electrons [with one node in the radial wavefunction located close to the C$_{60}$-radius $R=(\ri+\ro)/2=6.7$]. The free parameter $V_0=0.68$ is used to adjust the KS energy of the highest occupied molecular orbital (HOMO) to the ionization potential of C$_{60}$, $-\epsilon_\mathrm{HOMO}=\Ip\simeq 0.28$. The HOMO of our model is a $\pi$-orbital of angular momentum quantum number $\ell=4$.  Figure~\ref{figpotential} illustrates and summarizes the ground state configuration from which we start the time-dependent calculations. 

\subsection{Collective modes $\omegaMie$ and $\omegap$}
\begin{figure}
\includegraphics[width=0.4\textwidth]{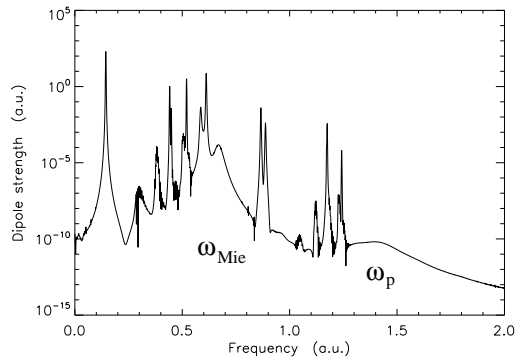}
\caption{Dipole response of the C$_{60}$ model system. Narrow lines (single-particle transitions) on top of two broad structures (surface or Mie plasmon $\omegaMie\simeq 0.7$ and volume plasmon $\omegap\simeq 1.4$) are observed. The Mie plasmon corresponds to homogeneous dipole-like oscillations of the electron density with respect to the ions. The volume plasmon (in general a breathing mode)  is visible in our dipole spectra since it contains a nonvanishing dipole component. The dipole strength is normalized such that its integral equals $N=250$.
\label{linresp}}
\end{figure}

In order to characterize the collective response of the model C$_{60}$ we apply the real-time method proposed in Ref.~\cite{yabana}. To that end we solve the time-dependent KS (TDKS) equation \cite{koval}
\beq \imagi\pabl{}{t} \ket{\psi_j(t)} = [T+V+\VI(t) +\VH + \Vxc]  \ket{\psi_j(t)} \label{tdkseq}\eeq
with 
\beq \VI(t)=\vektA(t)\cdot\vektp\eeq 
where $\vektA(t)=\Ahat \vekte_z \Theta(t)$ is a vector potential describing a $\delta$-like electric field $\vektE(t)=-\partial\vektA/\partial t = \Ahat \delta(t)$ in dipole approximation. 
From the Fourier-transform of the dipole 
\beq d_z(t) = \int\diff^3 r\ z\, n(\vektr,t) \eeq 
the spectrum $S(\omega) = \vert d_z(\omega) \vert^2$
is calculated. Figure~\ref{linresp} shows that the linear dipole response consists of several narrow lines (single-particle transitions) that sit on top of two broad structures (the surface and volume plasmon, respectively). Closer inspection shows that transitions of the type $\sigma\ell \to \pi(\ell\pm 1)$, $\pi\ell \to \sigma(\ell\mp1)$ contribute to the surface (or Mie) plasmon  $\omegaMie$ and  transitions between $\sigma$-states and (initially unoccupied) $\delta$-states (with two radial nodes) to the volume plasmon $\omegap$.

\section{Results} \label{resu}
In this Section we shall present and discuss our results for dipole spectra $S(\omega)$ of our model C$_{60}$ when exposed to Gaussian and trapezoidal laser pulses of various peak intensities and wavelengths.

\subsection{From linear to nonlinear plasmon excitation} \label{lintononlin}
We solved the TDKS equation \reff{tdkseq} for Gaussian pulses with a  vector potential of the form 
\beq A(t) = -\frac{\Ehat}{\omegalaser} \sin(\omegalaser t)\ \eulere^{-a^2(t)} \eeq
with
\beq a(t) = 3 \left[ \frac{\omegalaser t}{n \pi} -1\right]. \eeq
There are $0.278\,n$ cycles within the FWHM of the Gaussian pulse (with respect to the electric field or the vector potential) centered around $t=(2\pi/\omegalaser)(n/2)$.
We started the simulation from the ground state at $t=0$ and stopped at $t=(2\pi/\omegalaser) n$ with $n=8$.

\begin{figure}
\includegraphics[width=0.4\textwidth]{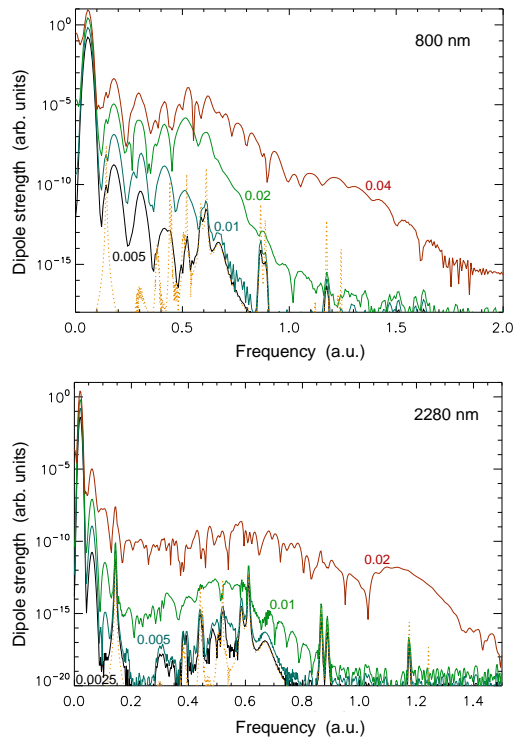}
\caption{(color online). Harmonic spectra of the  C$_{60}$ model for Gaussian laser pulses with  $\omegalaser=0.057$  ($\lambda=800$\,nm, upper panel) and  $\omegalaser=0.02$  ($\lambda=2280$\,nm, lower panel). The values of $\Ehat$ are given in the plots. The linear response profile from Fig.~\ref{linresp} is included (dotted). The field amplitude $\Ehat=0.04$ corresponds to the intensity $5.6\cdot 10^{13}$W/cm$^2$.
\label{lintononlinfig}
}
\end{figure}

Figure \ref{lintononlinfig} shows the transition from the linear to the nonlinear regime. At very low field amplitude ($\Ehat=0.0025$ and $0.005$ at 2280 and 800\,nm, respectively) the dipole spectra display replicas of the linear response profile on a very low level, depending on the bandwidth of the applied laser pulse. Upon doubling the field amplitude ($\Ehat=0.005$ and $0.01$ at 2280 and 800\,nm, respectively) the signal in the dipole spectrum is quadrupled, as expected in the linear regime. The corresponding values of intensity are $0.9\cdot 10^{12}$W/cm$^2$ and $3.5\cdot 10^{12}$W/cm$^2$. However, with further increasing laser intensity, plateaus develop and the high harmonic-signal increases rapidly over a wide frequency range. One may argue that this increase of the harmonic signal is just due to the standard harmonic generation mechanism while the collective response is still within the linear regime and thus not visible at higher laser intensities. The next Subsection is hence devoted to identify plasmon {\em enhancements} and their wavelength dependence.

\subsection{Plasmon enhancements and wavelength dependence} \label{wavelengthdep}
\begin{figure}
\includegraphics[width=0.4\textwidth]{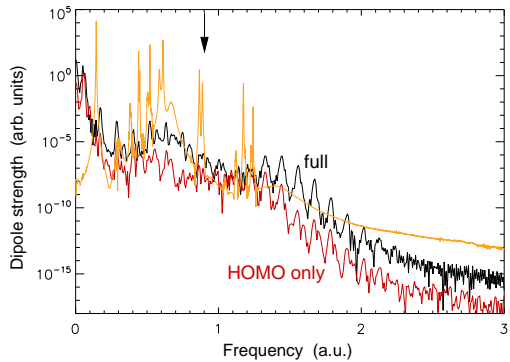}
\caption{(color online). Harmonic spectra of the  C$_{60}$ model for $\Ehat= 0.05$, $\omegalaser=0.057$  ($\lambda=800$\,nm), and an 8-cycle trapezoidal laser pulse with 2-cycles up and down ramps \cite{pulseshaperemark}. The full spectrum and the one just from the valence KS electron ('HOMO only') are shown. The linear dipole response from Fig.~\ref{linresp} is included (shifted vertically). The vertical arrow indicates the standard cut-off $3.17\Up+\vert\epsilon_\mathrm{HOMO}\vert $.  
\label{w0-057spec}
}
\end{figure}

Figure~\ref{w0-057spec} shows the harmonic spectra $S(\omega)$ as calculated from the full dipole and the outermost orbital density only ('HOMO only') for an 8-cycle, (2,4,2) trapezoidal 800-nm laser pulse, i.e.,  with 2-cycles up and down ramps and 4 cycles of constant amplitude $\Ehat=0.05$ \cite{pulseshaperemark}.   The difference between the two harmonic spectra clearly indicates that not just the valence electron contributes to the emission. Enhancements by two orders of magnitude around frequencies at which the system displays collective modes are visible. The standard cut-off known from atomic HOHG is at $3.17\Up+\vert\epsilon_\mathrm{HOMO}\vert $ (with $\Up=\Ehat^2/(4\omegalaser^2)$ the ponderomotive energy) and indicated by an arrow. The real cut-off, however, is extended to higher harmonic frequencies because recombination into orbitals with higher ionization potentials $\vert\epsilon_j\vert > \vert\epsilon_\mathrm{HOMO}\vert$ takes place. Note that the latter is possible without violation of the Pauli principle (unless KS electrons are frozen in the respective states). An extension of the standard harmonic plateau in a multielectron system---presumably of the same origin---has also been observed in Ref.~\cite{zanghell}.

In the following we show that with increasing laser wavelength the emission spectra become more and more SAE-like in the sense that all collective response is less efficient than the standard harmonic generation by the outermost electron at the respective frequency. In the SAE calculations we also start from the DFT groundstate but freeze the potentials $\VH$ and $\Vxc$ for the propagation of the valence KS orbital. 

\begin{figure}
\includegraphics[width=0.4\textwidth]{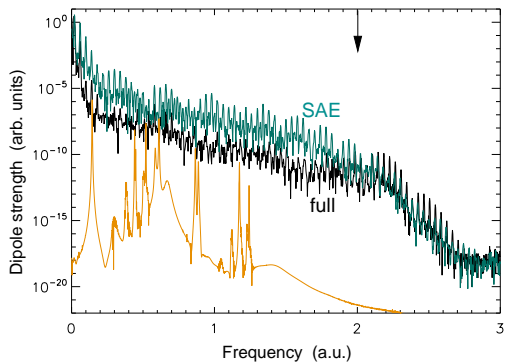}
\caption{(color online). Emission for  $\omegalaser=0.02$  ($\lambda=2280$\,nm) and  $\Ehat=0.03$ (other parameters as in Fig.~\ref{w0-057spec}). The results from a full TDKS calculation ('full') and a SAE-simulation are shown. The linear dipole response from Fig.~\ref{linresp} is included (shifted vertically).  The vertical arrow indicates the standard cut-off. 
\label{w0-02spec}
}
\end{figure}

Figure~\ref{w0-02spec} shows that at $\lambda=2280$\,nm there are still substantial differences between the SAE-result and the full TDKS calculation. First, the SAE yield is higher because the ionization step in the three step scenario described above is more efficient for a frozen potential since there is no polarization which counteracts the laser field. Second, the plasmon emission included in the full result obscures the oscillatory structure from which structural information (i.e., in our case the  C$_{60}$ radius and the width of the spherical jellium shell) could be obtained. Only in the (extended) cut-off region full and SAE-result agree very well because there are no collective modes at such high frequencies. 

\begin{figure}
\includegraphics[width=0.4\textwidth]{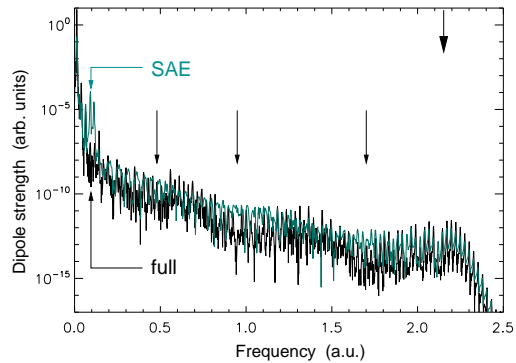}
\caption{(color online). Emission for  $\omegalaser=0.013$  ($\lambda=3508$\,nm) and  $\Ehat=0.02$. The results from a full time-dependent KS calculation ('full') and a SAE-simulation are shown.  The bold vertical arrow indicates the standard cut-off, the three thin vertical arrows local minima in the envelope of the full spectrum. 
\label{w0-013spec}
}
\end{figure}

At the even longer wavelength $\lambda=3508$\,nm the full TDKS result agrees well with the SAE result, as is shown in Fig.~\ref{w0-013spec}. Also the cut-off is at the expected position, indicating that recombination into states with orbital energies $\vert\epsilon\vert > \vert \epsilon_\mathrm{HOMO}\vert$ is insignificant. A closer inspection of the individual response of all the KS electrons shows that the standard HOHG generation of the HOMO KS electrons (i.e., the two spin-degenerate ones with $\ell=4$ and $m=0$) clearly dominates. Hence, long wavelengths are advantageous for imaging schemes which are based on interference structures in the HOHG spectra predicted by strong field-theoretical treatments \cite{lein05} in SAE approximation. 
However, the efficiency of HOHG also decreases with increasing laser wavelength \cite{schiessl}. The fact that the efficiency of the collective response decreases even faster is one of the main results of this work. 

\subsection{Identifying the mechanism}
In Sec.~\ref{lintononlin} we showed that at low field strengths the collective response increases linearly with the field strength (i.e., the signal in the dipole spectra quadratically) while at higher intensities the standard high harmonic plateau develops, which is an entirely nonlinear phenomenon. In Sec.~\ref{wavelengthdep} we showed that plasmon enhancements are present although they decrease relative to the standard high harmonic plateau with increasing wavelength. This means that there must be some nonlinear effect at work which is able to generate a collective response of comparable strength as the standard high harmonics. The latter are due to returning electrons which recombine. The obvious guess is to attribute the collective response also to the returning electrons so that the similar efficiency of harmonic emission via the SAE and via the collective mechanism can be understood  if recombination with emission of a photon and with excitation of a plasmon (followed by emission of a photon) are similarly efficient. In this Subsection we support the viewpoint that the recolliding electrons indeed excite collective modes  by analyzing our numerical results in more detail.

In our TDDFT simulations we use a spherically symmetric  imaginary potential $W(r)= -\imagi W_0 (r/\Rg)^{16}$ with $W_0=100$ and $\Rg$ the radius of the numerical grid. The imaginary potential serves as an absorber of probability density approaching the boundary of the numerical grid \cite{koval}. Usually the grid is chosen big enough so that only the probability density corresponding to never-returning electrons is absorbed and thus the imaginary potential does not affect the relevant dynamics taking place in the interior of the numerical grid where $W(r)$ is negligible. However, in order to test whether recolliding electrons are responsible for {\em both} the standard harmonic generation {\em and} the plasmon enhancements, we may absorb probability density representing electrons of a certain excursion amplitude $\zhat$ by moving the imaginary potential closer to the C$_{60}$. If the plasmon enhancements are due to recolliding electrons we then expect the harmonic signal and the plasmon signal to drop. If, instead, the plasmon enhancements are due to some other yet unknown nonlinear effect which does not require {\em returning} electrons, then the  harmonic signal should drop while the plasmon signal sustains. 

\begin{figure}
\includegraphics[width=0.4\textwidth]{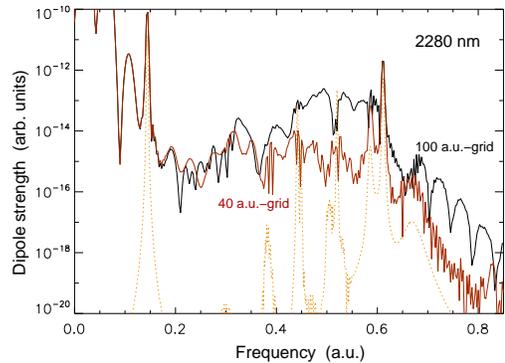}
\caption{(color online). Same as in Fig.~\ref{lintononlinfig}, lower panel, for $\Ehat=0.01$ but two different grid sizes (indicated in the plot). The linear response profile from Fig.~\ref{linresp} is included again (dotted).
\label{absbound}
}
\end{figure}

\begin{figure}
\includegraphics[width=0.4\textwidth]{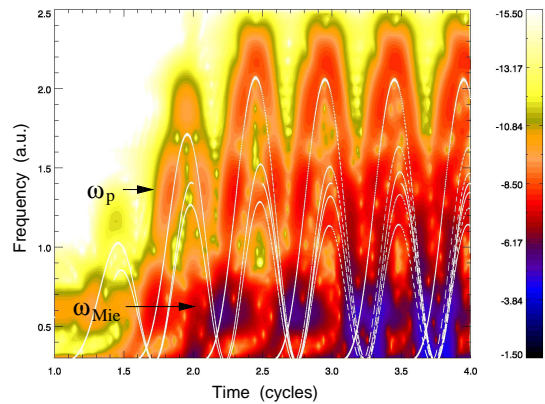}
\caption{(color). Logarithmically scaled contour plot of the time-frequency analyzed dipole emission $\log_{10} \vert d_z(t,\omega)\vert^2$ for the parameters of Fig.~\ref{w0-02spec}. The white lines indicate the classical solutions of returning electrons (see text). The positions of the Mie surface plasmon and the volume plasmon are indicated.
\label{timefrequ}
}
\end{figure}

Figure~\ref{absbound} shows dipole spectra for $\lambda=2280$\,nm and $\Ehat=0.01$ (i.e., the second highest intensity shown in the lower panel of Fig.~\ref{linresp}) for two grid sizes. The excursion amplitude of a free electron in this case is $\zhat=\Ehat/\omegalaser^2=25$. Hence we expect the $\Rg=100$-grid to comprise all the relevant electron dynamics whereas on the $\Rg=40$-grid some electrons will be already inhibited from returning to the C$_{60}$ because the corresponding probability density is absorbed. In fact, Fig.~\ref{absbound} shows that parts of the plateau are removed in the spectrum for the smaller grid. Only the single particle transition lines close to $\omega=0.6$ are unaffected by the absorbing boundary, showing that these transitions are not excited by recolliding electrons but---presumably---by multiphoton resonances. However, besides these resonant transitions the whole plateau is suppressed. We thus conclude that the returning electrons are essential for the excitation of the collective modes.  
This conclusion is further supported by a time-frequency analysis of the dipole $d_z(t)$. To that end a spectral window is applied to $d_z(\omega)$. The result is transformed back, which corresponds to the spectral filtering of certain harmonics for the generation of attosecond pulses in experiments \cite{ago04}. The result is shown in Fig.~\ref{timefrequ}. The emission follows overall nicely the classical ``simple man's theory'': the classical return-times of electrons with return-energy $\energy_\mathrm{ret}$ (which contribute to the emission of harmonic radiation at a frequency $\omega=\energy_\mathrm{ret} + \vert\epsilon_\mathrm{HOMO}\vert$) are indicated by white trajectories in the frequency-time plane. It is seen that the plasmon emission is correlated with the return of electrons. Whenever there are recolliding electrons having the right energy to excite a plasmon, enhanced emission is observed. Due to the large width of the collective resonances the emission decays before the next returning electron collides.

\section{SAE Lewenstein model {\em vs} long-wavelength TDDFT-result } \label{lewe}
We now show that the structure in the HOHG spectrum of Fig.~\ref{w0-013spec} is indeed similar to what one expects from the strong field approximation applied to HOHG, i.e., the so-called Lewenstein-model \cite{lewen}. Within the Lewenstein-model the dipole expectation value for an infinite, linearly polarized laser pulse

\beq \vektE(t)=\Ehat\vekte_z \cos\omegalaser t, \qquad E(t)=-\partial_t A(t)\eeq is given by 
\begin{eqnarray} d_z^\mathrm{(L)}(t) &=& \imagi\int_0^\infty\!\!\!\!\diff\tau \left(  \frac{2\pi}{\imagi\tau}\right)^{\!\!3/2}\!\!\! \mu_z^*[k(t,\tau)+A(t)] \label{lewenstein} \\
&&\qquad  \times  \exp [-\imagi S(t,\tau)]  \Ehat\cos[\omegalaser(t-\tau)]  \nonumber \\
&&\qquad \times \mu_z[k(t,\tau)+A(t-\tau)] + \mathrm{c.c.} \nonumber
\end{eqnarray}
where $\tau$ is the travel-time of the electron between ionization and recombination, 
\beq \mu_z(p_z)=\bra{p_z} z \ket{\Psi_0},\eeq 
$k(t,\tau)$ is the saddle-point momentum 
\beq k(t,\tau) = -\Ehat\ \frac{\cos\omegalaser t-\cos\omegalaser(t-\tau)}{\omegalaser^2\tau},\eeq 
and $S(t,\tau)$ the saddle-point action 
\begin{eqnarray} S(t,\tau) &=& (\Up-\epsilon_\mathrm{HOMO})\tau -\frac{2\Up(1-\cos\omegalaser\tau)}{\omegalaser^2\tau} \nonumber \\
&& - \frac{\Up C(\tau)\cos[(2t-\tau)\omegalaser]}{\omegalaser}
\end{eqnarray}
with 
\beq C(\tau)=\sin\omegalaser\tau-\frac{4}{\omegalaser\tau} \sin^2(\omegalaser\tau/2).\eeq 
For a derivation of \reff{lewenstein} the reader is referred to the original work in Ref.~\cite{lewen}.

The target-dependence of the HOHG spectra enters in \reff{lewenstein} via the initial state $\Psi_0$ through the ionization and recombination matrix elements $\mu_z[k(t,\tau)+A(t-\tau)]$ and $ \mu_z^*[k(t,\tau)+A(t)]$, respectively \cite{ciapp}. We assume an initial state of the form $\Psi_0(\vektr) = \Phi_0(r) Y_{\ell 0}(\theta,\varphi) /r$ with $ Y_{\ell m}(\theta,\varphi)$ a spherical harmonic and model the valence $\pi$-orbital using a radial wavefunction $ \Phi_0(r)/r = (2\Delta)^{-1/2} $ for $R-\Delta < r\leq R$,  $-(2\Delta)^{-1/2}$  for $R < r\leq R+\Delta$, and zero otherwise. Here,  $\Delta$ is half the thickness of the C$_{60}$-shell, i.e., $\Delta=(\ro-\ri)/2=1.4$. Assuming further $\vert p_z\Delta\vert\ll 1$ and, e.g., $\ell=0$ we obtain 
\beq \mu_z(p_z) \sim \frac{1}{p_z^2}(\sin p_zR - p_zR \cos p_zR +p_z^2R^2 \sin p_zR) \label{mu}\eeq 
and a similar but more lengthy expressions for $\ell=4$. One clearly sees that structural information (i.e., the C$_{60}$-radius $R$) is 'encoded' in $\mu_z(p_z)$. If the approximation $\vert p_z\Delta\vert\ll 1$ is not made, also information about the shell thickness $2 \Delta$ is included in the matrix element $\mu_z(p_z)$.

\begin{figure}
\includegraphics[width=0.4\textwidth]{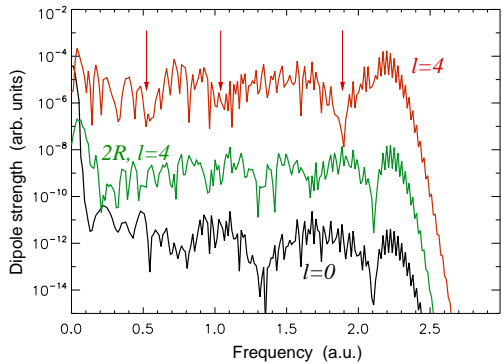}
\caption{(color online). Harmonic spectra $S(\omega)$, calculated from Eq.~\reff{lewenstein} for $\ell=4$, $\ell=0$, and $\ell=4$ but twice the radius $R$. The laser parameters are the same as in Fig.~\ref{w0-013spec} [$\omegalaser=0.013$  ($\lambda=3508$\,nm) and  $\Ehat=0.02$]. $R=6.7$ and $\Delta=1.4$ was used. The three thin vertical arrows indicate local minima in the envelope of the spectrum for $\ell=4$. 
\label{lewenplot}
}
\end{figure}

Figure~\ref{lewenplot} shows the harmonic spectra obtained from the Fourier-transform $d_z^\mathrm{(L)}(\omega)$ of  Eq.~\reff{lewenstein} for $\ell=4$ and $\ell=0$ and the laser parameters of Fig.~\ref{w0-013spec}. The positions of the minima in the envelope of the HOHG spectra depend on the initial $\ell$ quantum number and the C$_{60}$ radius $R$. In order to illustrate this dependency the spectra for $\ell=0$, $\ell=4$, and $\ell=4$ but with  the radius doubled are shown. The minima indicated in the $\ell=4$-spectrum by vertical arrows may be compared with those of the time-dependent DFT  result in  Fig.~\ref{w0-013spec}. The latter are at $\omega\simeq 0.5$, $0.95$, and $1.7$. The arrows in Fig.~\ref{lewenplot} are at $\omega\simeq 0.52$, $1.05$, and $1.9$, which is in reasonable agreement. Note that the agreement would be worse if one attempted to compare  with the $\ell=0$-spectrum, let alone with the spectrum for $\ell=4$ and doubled radius, which is qualitatively different since there is at least one more pronounced minimum in the envelope.

\section{Analytical model for harmonic generation including recollision-induced collective excitations} \label{collmodel}
For systems with a single active electron, harmonic spectra are usually analyzed using the strong field approximation (or Lewenstein model) \cite{lewen,becker}, as we did in the previous section.
Emission into a mode with frequency $\omega$ and polarization $\vece_{\lambda}$, $\lambda=1,2$, by a system with only a single active electron can also be described by the amplitude
\begin{widetext}
\beq
M_{\rm SAE}(\omega,\lambda)=-\int \!\!\diff^3p\,\int_{-\infty}^{\infty}\!\!\diff t\,
\langle \Psi_0,1_{\omega,\lambda}|{\hat V}_{\rm rad}^{+}|0_{\omega,\lambda},\vecp\rangle
\int_{-\infty}^{t}\!\!\diff t'\, \langle \vecp|{\hat V}|\Psi_0\rangle.
\label{single}
\eeq
Here, $|\Psi_0\rangle$ is the single-electron ground state, $|\vecp\rangle$ is the Volkov state of drift momentum $\vecp$, $|n_{\omega,\lambda}\rangle$ is the Fock state of the harmonic radiation field with $n$ photons in the respective mode ($n=0,1$ in our case) and ${\hat V}_{\rm rad}^{+}$ and ${\hat V}$ are the interaction operators coupling to the radiation and the laser field, respectively,
\beq
{\hat V}_{\rm rad}^+=-\imagi\sqrt{\frac{2\pi\omega}{{\cal V}}}\vecr\cdot\vece_{\lambda}{\hat a}^+_{\omega,\lambda}\ \eulere^{\imagi\omega t},
\eeq
\beq
V =E(t)z,
\eeq
with ${\hat a}^+$ the photon creation operator and ${\cal V}$ the quantization volume.  The amplitude \reff{single} describes an electron which is lifted from the ground state to a Volkov state by the laser field at time $t'$ and emits a harmonic photon upon recombination at time $t>t'$.
The harmonic spectrum is given by the square modulus of (\ref{single}) and appears to be virtually identical to the spectrum found from the dipole (\ref{lewenstein}) \cite{becker}.
In the dipole approximation we use here the wavevector $\vecK$ of the emitted photon does not appear in the amplitude (\ref{single}).

Now we introduce a similar amplitude which accounts for the collective modes:
in addition to the pathway described by (\ref{single}) the recombining electron may excite collective modes which then relax upon emission of a harmonic photon.
The amplitude for such a process reads
\beq
M_{\rm coll}(\omega,\lambda)=\sum_j\sum_L\int_{-\infty}^{\infty}\!\!\diff t\,
\langle 0_j,1_{\omega,\lambda}|{\hat V}_{\rm rad}^{+}|0_{\omega,\lambda},L_j\rangle\int \!\!\diff^3p\, \int_{-\infty}^{t}\!\!\diff t'\,\langle \Psi_0,L_j|{\hat U}|0_j,\vecp\rangle
\int_{-\infty}^{t'}\!\!\diff t''\, \langle \vecp|{\hat V}|\Psi_0\rangle.
\label{collective}
\eeq
\end{widetext}
Here, $|0_j\rangle$ and $|L_j\rangle$ are the ground and the $L$-th excited state of a collective mode, labelled by $j$ (e.g., surface or volume oscillations).
The interaction energy between the electron and the residual electron cloud is described by the operator ${\hat U}$.
The amplitude (\ref{collective}) is a straightforward generalization of the Lewenstein model to the case when collective modes can be involved in the emission process.

To evaluate the collective amplitude (\ref{collective}) a certain model for the description of the collective modes and their interaction ${\hat U}$ with the active electron is required.
In order to estimate the relative contribution of the SAE and the collective pathways to the radiation spectrum we use a simple model which takes collective degrees of freedom into account as two noninteracting harmonic oscillators with eigenfrequencies $\omega_{\rm Mie}\simeq 0.7$ and $\omegap\simeq 1.4$ (i.e., the surface and the volume plasmon in the C$_{60}$-model above).
The respective widths of the plasmons are taken as $\Gamma_{\rm Mie(p)}\approx 0.2$.
The main physical mechanism which generates these widths is a coupling between collective and single-electron degrees of freedom.
This can also be interpreted as collisionless or Landau damping of collective modes in a finite system \cite{zar,kreibig}.
To obtain an explicit form for the interaction operator ${\hat U}$ we employ a rigid sphere model (RSM) in which the electron cloud is treated as an incompressible homogeneous sphere which may oscillate around its equilibrium position.
%The ions are also distributed homogeneously over a sphere of the same radius $R$.
Note that on the level of modelling in this Section it does not matter whether we consider a homogeneous sphere or a spherical shell.
Within the RSM the interaction operator has the form
\beq
U(\vecr,\vecX)=\frac{(N-1)}{R}
\left\{\begin{array}{l}\displaystyle \frac{3}{2}-\frac{(\vecr-\vecX)^2}{2R^2}, \ |\vecr-\vecX|\le R,\\
\displaystyle \frac{R}{|\vecr-\vecX|}, \ \ \ \ \ \  \ \ \ |\vecr-\vecX|> R,\end{array}
\right.
\label{U}
\eeq
where $N=240$ is the number of electrons, $\vecr$ is the active electron's position and $\vecX$ is the center-of-mass displacement of the electron cloud.
Because of the relatively high energies of the plasmons only the first excited collective states are relevant in the sum over $L$ in (\ref{collective}).
For a first excited state $X\simeq1/\sqrt{(N-1)\omega_{\rm Mie(p)}}\ll R$ so that with high accuracy (e.g., taking $N=240$ and $\omegaMie=0.7$ one estimates $X\simeq 0.08$) one may simplify (\ref{U}) keeping only the linear term with respect to the center-of-mass displacement $\vecX$:
\beq
U(\vecr,\vecX)\simeq U_0(r)+\frac{N-1}{R^3}\vecr\cdot\vecX
\left\{\begin{array}{l}\displaystyle 1,~~~~~~~~~r\le R,\\
\displaystyle R^2/r^3,~~~r> R\end{array}.
\right.
\label{Usimpl}
\eeq
Next, we assume that the electron excursion amplitude in the laser field ${\hat z}=\Ehat/\omegalaser^2$ is less than or comparable to the cluster size $R$.
Then, with reasonable accuracy, we may use
\beq
U(\vecr,\vecX)\simeq \frac{N-1}{R^3}\,\vecr\cdot\vecX
\label{Usimpl2}
\eeq
instead of (\ref{Usimpl}).
Within this approximation an explicit relation between the amplitudes (\ref{collective}) and (\ref{single}) can be derived. To this end we first evaluate the emission matrix element in (\ref{single}),
\beq
\langle \Psi_0,1_{\omega,\lambda}|{\hat V}_{\rm rad}^{+}|0_{\omega,\lambda},\vecp\rangle=
-\imagi \sqrt{\frac{2\pi\omega}{{\cal V}}}\langle\Psi_0|\vecr\cdot\vece_{\lambda}|\vecp\rangle\ \eulere^{\imagi\omega t}.
\label{em1}
\eeq
%Here we used the standard relation $\langle a|\vecp|b\rangle=-\imagi m\omega_{ab}\langle a|\vecr|b\rangle$ which is only correct if the states $|a\rangle$ and $|b\rangle$ are eigenstates of the same Hamiltonian.
%Within our model this is not strictly the case for the states $|g\rangle$ and $|\vecp\rangle$. In deriving \reff{em1} we have further taken into account that in a linearly polarized field $\vecA=A\vece_0$, after integration over all momenta $\vecp$ in (\ref{single}), the unit vector $\vece_0$ along the polarization direction should appear so that we may replace $\vecr\cdot\vece_{\vecK,\lambda}$ by the product $(\vece_0\cdot \vece_{\vecK,\lambda})  (\vecr\cdot \vece_0)$.
A similar procedure for the emission matrix element in (\ref{collective}) yields
\begin{eqnarray}
\lefteqn{\langle 0_{\rm Mie(p)},1_{\omega,\lambda}|{\hat V}_{\rm rad}^{+}|0_{\omega,\lambda},1_{\rm Mie(p)}\rangle}\label{em2} \\
&=& -\imagi\sqrt{\frac{\pi(N-1)\omega}{{\cal V}\omega_{\rm Mie(p)}}}\vece_z\cdot\vece_{\lambda}\ \eulere^{\imagi(\omega-\omega_{\rm Mie(p)})t-\Gamma_{\rm Mie(p)}t/2}. \nonumber
\end{eqnarray}
Here we used the fact that for the harmonic oscillator $\langle 0|z|1\rangle=1/\sqrt{2M\Omega}$ with $M=(N-1)m$ and $\Omega=\omega_{\rm Mie(p)}$ in our case.
Also we take into account that the oscillator is exited along the polarization direction given by the unit vector $\vece_z$.

Rearranging the time-integrations, the amplitude (\ref{collective}) can be also written as
\begin{widetext}
\beq
M_{\rm coll}(\omega,\lambda)=\sum_j\sum_{\rm Mie,p}
\int \!\!\diff^3p\,\int_{-\infty}^{\infty}\!\!\diff t'\,\langle\Psi_0,1_{\rm Mie(p)}|{\hat U}|0_{\rm Mie(p)},\vecp\rangle
\int_{t'}^{\infty}\!\!\diff t\,
\langle 0_{\rm Mie(p)},1_{\omega,\lambda}|{\hat V}_{\rm rad}^{+}|0_{\omega,\lambda},1_{\rm Mie(p)}\rangle
\int_{-\infty}^{t'}\!\!\diff t''\,\langle \vecp|{\hat V}|\Psi_0\rangle.
\label{collective2}
\eeq
Now the inner integral over $t$ can be evaluated explicitly using (\ref{em2}). The result reads
\beq
\int_{t^{\prime}}^{\infty}\!\!\diff t\,
\langle 0_{\rm Mie(p)},1_{\omega,\lambda}|{\hat V}_{\rm rad}^{+}|0_{\omega,\lambda},1_{\rm Mie(p)}\rangle
= -\imagi\sqrt{\frac{\pi(N-1)\omega}{{\cal V}\omega_{\rm Mie(p)}}}\frac{\vece_z\cdot\vece_{\lambda}\ \eulere^{\imagi(\omega-\omega_{\rm Mie(p)})t'}}{\Gamma_{\rm Mie(p)}/2-\imagi(\omega-\omega_{\rm Mie(p)})}. \label{em3}
\eeq
Finally, using the standard expression for the coordinate matrix element of the harmonic oscillator $\langle 0|z|1\rangle=1/\sqrt{2M\Omega}$ and (\ref{Usimpl2}) one obtains for the first matrix element in (\ref{collective2})
\beq
\langle\Psi_0,1_{\rm Mie(p)}|{\hat U}|0_{\rm Mie(p)},\vecp\rangle=\frac{1}{R^3}\sqrt{\frac{N-1}{2\omega_{\rm Mie(p)}}}
\langle\Psi_0|z|\vecp\rangle\ \eulere^{\imagi\omega_{\rm Mie(p)}t}.
\label{rec}
\eeq
Collecting Eqs.(\ref{em1})--(\ref{rec}), we may express the amplitude (\ref{collective}) via (\ref{single}) as
\beq
M_\mathrm{coll}=\imagi\frac{N-1}{2R^3}\left\{\frac{\omega_{\rm Mie}^{-1}}{\Gamma_{\rm Mie}/2-\imagi(\omega-\omega_{\rm Mie})}+
\frac{\omega_{\rm p}^{-1}}{\Gamma_{\rm p}/2-\imagi(\omega-\omega_{\rm p})}\right\}M_\mathrm{SAE}.
\label{ratio1}
\eeq
\end{widetext}

Equation (\ref{ratio1}) shows that collective modes may lead to enhancements in the HOHG spectrum around  the respective plasmon frequencies.
For the plasmon enhancements to be detectable  $\vert M_\mathrm{coll} \vert^2 > \vert M_\mathrm{SAE} \vert^2$ should hold. For the ratio of collective to SAE HOHG efficiency we obtain
\beq
\frac{|M_\mathrm{coll}|^2}{\vert M_\mathrm{SAE}(\omega=\omega_{\rm Mie(p)})\vert^2} \simeq \left\lbrack\frac{N-1}{R^3\omega_{\rm Mie(p)}\Gamma_{\rm Mie(p)}}\right\rbrack^2.
\label{ratio2}
\eeq
For $N=240$, $R=6.7$, $\Gamma_\mathrm{Mie}=\Gamma_\mathrm{p}=0.2\,$ the ratio (\ref{ratio2}) is above 10 for the surface and about unity for the volume plasmon.

The ratio (\ref{ratio2}) does not depend on the laser parameters anymore whereas in our TDDFT results we observe a wavelength-dependent relative efficiency of the plasmon enhancements. With increasing laser intensity or wavelength the electron's excursion amplitude is increasing and the approximation (\ref{Usimpl2}) for the interaction between the active electron and the electron cloud becomes invalid. Without the assumption of small excursion amplitudes (as compared to the cluster radius) a simple relationship of the type (\ref{ratio1}) cannot be established. Qualitatively it is quite obvious, however, that with increasing excursion amplitude distances $r\simeq R$ [for which (\ref{Usimpl}) is sizeable] contribute less and less to the spatial matrix element (\ref{rec}).
As a consequence  the standard single-electron HOHG spectrum dominates for $\Ehat/\omegalaser^2 \gg R$. In fact, $\Ehat/\omegalaser^2=15.4$, $75.0$, and $118.3$ in Figs.~\ref{w0-057spec},  \ref{w0-02spec}, and \ref{w0-013spec}, respectively, supporting our statement.

The results (\ref{ratio1}) and (\ref{ratio2}) were derived making several approximations besides the one of small excursion amplitudes.
For example, the surface and the volume plasmons were treated as independent.
This makes sense if they are well separated from each other, i.e., $|\omegaMie-\omegap|\gg (\Gamma_{\rm Mie}+\Gamma_{\rm p})/2$, which is actually not fulfilled in the case of C$_{60}$.
Another simplification was that we applied the RSM for the description of the electron cloud.
Within this model the volume plasmon simply does not exist. In a more realistic description one should use two different interaction potentials instead of (\ref{U}) alone, which will lead to two different coefficients in (\ref{ratio2}). 

\section{Conclusions} \label{concl}
In conclusion, we predict a new recollision effect in the interaction of strong laser fields with multi-electron systems. Besides the usual high-order harmonic generation the recolliding electron may excite collective modes instead of emitting its energy directly as a harmonic photon. Via the recollision mechanism collective modes can be excited even if the incident laser is far off-resonant with the plasmon frequencies.  Using time-dependent density functional theory we have studied the wavelength-dependence of the process in the case of C$_{60}$. With increasing laser wavelength the dynamics becomes more and more single active electron-like. Experiments employing imaging techniques based on recolliding electrons are hence more likely to reveal clean structural information if sufficiently long wavelengths are used.

\section*{Acknowledgments}
This work was supported by the Deutsche Forschungsgemeinschaft and the Russian Foundation for Basic Research (S.V.P.).

%%%%%%%%%%%%%%%%%%%%%


\begin{thebibliography}{}
\bibitem{milo06} D.B.\ Milo\v{s}evi\'c, G.G.\ Paulus, D.\ Bauer, and W.\ Becker, J.\ Phys.\ B: At.\ Mol.\ Opt.\ Phys.\ {\bf 39}, R203 (2006).
\bibitem{ago04} P.\ Agostini and L.F.\ DiMauro, Rep.\ Prog.\ Phys.\ {\bf 67}, 813 (2004).
\bibitem{abecker05} A.\ Becker and F.H.M.\ Faisal, J.\ Phys.\ B: At.\ Mol.\ Opt.\ Phys.\ {\bf 38}, R1 (2005).
\bibitem{itani04} J.\ Itatani, J.\ Levesque, D.\ Zeidler, Hiromichi Niikura, H.\ P\'epin, J.C.\ Kieffer, P.B.\ Corkum, D.M.\ Villeneuve, Nature  {\bf 432}, 867 (2004).
\bibitem{lein07} M.\ Lein, J.\ Phys.\ B: At.\ Mol.\ Opt.\ Phys.\ {\bf 40}, R135 (2007).
\bibitem{patch07} Serguei Patchkovskii, Zengxiu Zhao, Thomas Brabec, and D.M.\ Villeneuve, \prl {\bf 97}, 123003 (2006); J.\ Chem.\ Phys.\ {\bf 126}, 114306 (2007).
\bibitem{hertelreview} I.V.\ Hertel, T.\ Laarmann, and C.P.\ Schulz, Adv.\ At.\ Mol.\ Opt.\ Phys.\ {\bf 50}, 219 2005.
\bibitem{c60exp} Ihar Shchatsinin, Tim Laarmann, Gero Stibenz, G\"unter Steinmeyer, Andrei Stalmashonak, Nick Zhavoronkov, Claus Peter Schulz, and Ingolf V.\ Hertel, J.\ Chem.\ Phys.\ {\bf 125}, 194320 (2006). 
\bibitem{puska} M.J.\ Puska and R.M.\ Nieminen, \pra {\bf 47}, 1181 (1993). 
\bibitem{bauerC60} D.\ Bauer, F.\ Ceccherini, A.\ Macchi, and F.\ Cornolti, \pra {\bf 64}, 063203 (2001).
\bibitem{yabana} K.\ Yabana and G.F.\ Bertsch, \prb {\bf 54}, 4484 (1996).
\bibitem{koval} D.\ Bauer  and P.\ Koval, Comp.\ Phys.\ Comm.\ {\bf 174}, 396 (2006).
\bibitem{pulseshaperemark} Trapezoidal pulses have the advantages (i) that the comparison with the semi-analytical results for constant amplitude pulses of Sec.~\ref{lewe} is more straightforward, and (ii) that cut-off positions do less sensitively depend on the carrier-envelope phase since $\Ehat$ is constant over several cycles (see Ref.~\cite{milo06} for a discussion of  carrier-envelope phase-effects). We checked that all our findings are qualitatively insensitive to the pulse shape.
\bibitem{zanghell} J.\ Zanghellini, Ch.\ Jungreuthmayer, and T.\ Brabec, \jpb {\bf 39}, 709 (2006).
\bibitem{lein05} M.\ Lein, \prl {\bf 94}, 053004 (2005).
\bibitem{schiessl}  K.\ Schiessl, K.L.\ Ishikawa, E.\ Persson, and J.\ Burgd\"orfer, \prl {\bf 99}, 253903 (2007).
\bibitem{lewen} M.\ Lewenstein, Ph.\ Balcou, M.Yu.\ Ivanov, A.\ L'Huillier, and P.B.\ Corkum, \pra {\bf 49}, 2117 (1994).
\bibitem{ciapp} M.F.\ Ciappina, A.\ Becker, and A.\ Jaro\'n-Becker, \pra {\bf 76}, 063406 (2007).
\bibitem{becker} W.\ Becker, S.\ Long, and J.K.\ McIver, \pra {\bf 50}, 1540 (1994).
\bibitem{zar} Yu.A.\ Malov and D.F.\ Zaretsky, Phys. Lett. A {\bf 177}, 379 (1993).
\bibitem{kreibig} U.\ Kreibig and M.\ Vollmer, Optical Properties of Metal Clusters (Springer, Berlin, 1995).

%\bibitem{salieres} P.\ Sali\`eres {\em et al.}, Science {\bf 292}, 902 (2001).
%\bibitem{becker} W.\ Becker {\em et al.}, Adv.\ At.\ Mol.\ Opt.\ Phys.\ {\bf 48}, 35 (2002).
%\bibitem{phstab} D.J.\ Jones {\em et al.}, Science {\bf 288}, 635 (2000); A.\ Apolonski {\em et al.}, \prl {\bf 85}, 740 (2000); A.\ Baltu\v{s}ka \etal, \prl {\bf 88}, 133901 (2001).
%\bibitem{fewcycexp} A.\ Baltu\v{s}ka \etal, Nature (London) {\bf 421}, 611 (2003); G.G.\ Paulus \etal, \prl {\bf 91}, 253004 (2003); H.\ Niikura \etal, Nature (London) {\bf 417}, 917 (2002); R.\ Kienberger \etal, Nature (London) {\bf 427}, 817 (2004). 
%\bibitem{phaseinfluence} D.B.\ Milo\v{s}evi\'c \etal, \prl {\bf 89}, 153001 (2002); Opt.\ Express {\bf 11}, 1418 (2003); Laser Phys.\ Lett.\ {\bf 1}, 93 (2004); S.\ Chelkowski \etal, Opt.\ Lett.\ {\bf 29}, 1557 (2004); \pra {\bf 70}, 013815 (2004).
%\bibitem{schafer} K.J.\ Schafer and K.C.\ Kulander, \pra {\bf 42}, 5794 (1990).
\end{thebibliography}
\end{document}